\documentstyle[12pt,epsfig]{article}
\def\be{\begin{equation}}
\def\ee{\end{equation}}

\title{Structures and intermittency in a passive scalar model}
\author{M.~Vergassola$^{1}$ and A.~Mazzino$^{2}$\\
\small{$^{1}$ CNRS, Observatoire de Nice, B.P. 4229,
06304 Nice Cedex 4, France.}\\
\small$^2$ INFM -
Dipartimento di Fisica, Universit\`a di Genova, I--16146
Genova, Italy.}
\begin{document}
\maketitle
\date{}
\begin{abstract}

A one-dimensional white-in-time passive scalar model is introduced.
Strong and persistent structures are shown to be present.
A perturbative expansion for the scaling exponents 
is performed around a Gaussian limit of the model. The resulting
predictions are compared with numerical simulations.

\end{abstract}

PACS number(s)\,: 47.10.+g, 47.27.-i, 05.40.+j
\vspace{5mm}\\
It is commonly observed 
in turbulent flows that coherent and well localized structures 
emerge in the sea of disorganized, random regions.
Well-known examples are provided by 2D and 3D Navier-Stokes turbulence\,:
numerical simulations show that structures appear respectively
in the form of quasi-circular vortices and thin, elongated vortex tubes.
No ab initio theory for the statistical properties of the
flows (and a fortiori of their dependence on the structures) 
is however available. A systematic statistical theory has recently
been developed for white-in-time passive field models. Their crucial 
property is that simultaneous correlation functions obey closed 
equations of motion \cite{RHK}.
Anomalous scaling and intermittency 
have in particular been considered. The general
mechanism for the appearance of anomalous scaling is associated with zero
modes of the equations for the correlation functions \cite{CL,GK,ShSi}.
For Kraichnan passive scalar model, anomalies first appear at the
level of fourth-order correlations [1\,-\,3, 5]. 
When a large-scale gradient is imposed,
also the third-order moment is anomalous \cite{AP}. 
Scaling exponents have been calculated by using perturbative 
expansions around Gaussian limits \cite{CL,GK}.
For magnetic fields a nonperturbative anomalous solution for
second-order correlations has been found \cite{MV}.
The zero mode responsible
for anomalous scaling comes from the balance in the inertial 
range between field lines stretching and eddy-damping. This suggests that
also for white-in-time models, 
structures could be present and play an important role. 
A crucial point is whether the perturbative expansions around Gaussian limits
are capable, as those in critical phenomena, 
to correctly capture the global statistical
effects of structures. Comparing the perturbative predictions with the
results of numerical simulations would clearly help clarifying this issue.
The $\delta$-correlation in time of the advecting velocity makes 
however simulations quite intricate \cite{KYC}.  
We have therefore concentrated our attention on the 
1D white-in-time passive scalar model presented here. 
As in Kraichnan model, the velocity is Gaussian and $\delta$-correlated in
time. Its
structure function scales with a positive exponent $\xi$ and the dynamics
has a Gaussian limit for $\xi\to 0$.

The simulations show that the activity of the scalar
is indeed concentrated in strongly localized peaks. 
Despite of the $\delta$-correlation in time of the advecting velocity, 
their lifetime can be significant. The mechanism of formation of the peaks
is associated with the stretching due to velocity gradients. The
peaks are
responsible for the observed very high tails of the probability distribution
functions (p.d.f.). 

The $\delta$-correlation in time ensures that correlation functions 
satisfy closed equations of motion.
The behaviour of second-order correlations is obtained analytically and
their scaling is normal. 
Predictions for the scaling exponents of higher-order
correlations are derived using the perturbation scheme in $\xi$ proposed in
Ref.~\cite{GK}. Analytic predictions (in the form of  
Pad\'e approximants) for the fourth and sixth-order structure 
functions are compared for $\xi=0.5$ with numerical results. \\

The equation of the model is
\be
\label{model}
\partial_t \phi + \nabla\,\left( v\,\phi\right) = \kappa \Delta\phi + f .
\ee
Here, both the velocity $v$ and the injection $f$ are Gaussian and
$\delta$-correlated in time. The velocity correlation function is
\be
\label{velocity}
\langle v(x,t)\,v(x',t')\rangle = \delta(t-t')\,\left[D_0-S(|x-x'|)\right].
\ee
The structure function $S$ scales as $S(x-x')= D\, |x-x'|^\xi$
(and $0<\xi <2$)
in the range of scales between the ultraviolet and the infrared 
cutoffs $\Lambda_{UV}$ and $\Lambda_{IR}$ (the smallest and the largest
scales in our problem). The large-scale injection 
satisfies $\langle f(x,t)\,f(x',t')\rangle =
\delta(t-t')\,F_L(x-x')$ with its Fourier transform $\hat{F}_L$ 
concentrated around wave numbers $O(1/L)$.
It is assumed that $\hat{F}_L$ vanishes at $k=0$, i.e.
$\int_{-\infty}^{+\infty}F_L(x)\,dx=0$.
This condition is associated with the fact that $\phi$ is a
``gradient field''. The equation for the gradients in Kraichnan model
indeed coincides with (\ref{model}) when vector components are omitted.
The gradient nature of $\phi$ appears even more clearly from the solution
for the second-order correlations presented later. 
The homogeneous part of the equation for the integrated field
$\theta(x,t)=\int^x \phi(y,t)\,dy$
(scaling with positive exponents) is
\be
\label{integra}
\partial_t\theta + v\,\nabla\theta = \kappa \Delta \theta .
\ee
Note that for (\ref{integra}) the maximum principle holds.
The stretching process due to the gradients of $v$ operates
on the gradient field $\phi$. 
To fully resolve the structures 
we have then preferred to deal directly with (\ref{model}).
This equation has been integrated by using a pseudo-spectral code 
with periodic boundary conditions. At each time step a new realization of 
both $v$ and $f$ is generated. The multiplicative term in (\ref{model})
is numerically treated explicitly (\`a la Ito). 
The term differing in the formulations Ito 
and Stratonovich (the latter being
relevant for (\ref{model})) is taken into account \cite{KP}. 
The anomalous exponents 
are not expected to depend on the type of dissipation. 
For the measurement of the scaling exponents we have then  
used a $k^8$ hyperdissipation, but we have also checked 
that structures are present for normal dissipation.
The resolution is $N=2^{14}$.
The injection is concentrated on the first mode. The velocity
$v$ is simply generated in Fourier space.
\begin{figure}
\begin{center}
\mbox{\epsfig{file=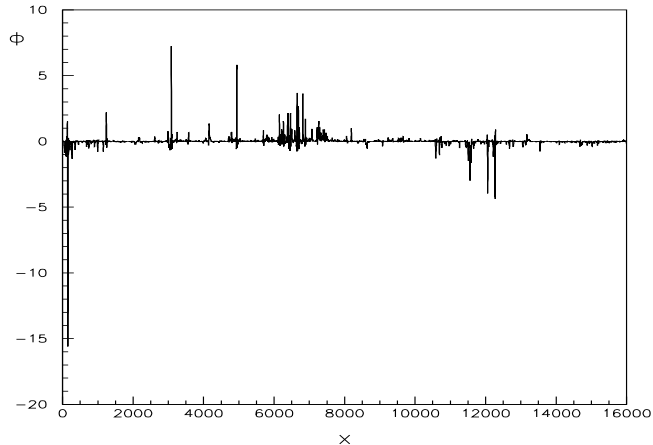,height=7cm,width=10cm}}
\end{center}
\vspace{-1cm}
\caption{The gradient field $\phi$ {\it vs} the spatial 
coordinate $x$ for $\xi=0.5$.} 
\label{fig1}
\end{figure}

A typical plot of the 
gradient field $\phi$ is shown in Fig.~1.
It is evident the presence of huge peaks of activity.
Their mechanism of formation (and persistence) 
can be immediately grasped from the original equation (\ref{model}). 
Neglecting diffusion and forcing, the extrema $\bar{\phi}$ 
of the field $\phi$ obey
$\partial_t \bar{\phi} = - \left(\nabla v\right)\bar{\phi}$.
Since $v$ is $\delta$-correlated in time, the logarithm of 
$\bar{\phi}$ 
evolves then as a Brownian motion. On the other hand,
it is known that random
walkers have a tendency not to change sign.
Specifically, let us start a walk from the
origin. The cumulative 
distribution for the fraction of the walk spent on the negative side
obeys the arc sine law \cite{Fell}\,: 
keeping the same sign for the whole walk and equipartition of the 
time between positive and negative values 
are the most and the least probable events, respectively.
In our case, this implies
that there will be very long periods of time when the 
stretching mechanism operates. Peaks such
as those in Fig.~1 can then be maintained for significant times, despite 
the $\delta$-correlation of the velocity.
This is indeed observed in 
our numerical simulations both for normal and hyper-dissipation. 
Strong peaks persist for long enough to make them clearly identifiable
and meaningful the use of the word ``structures'' to denote them.
The structures are evidently the 
cause of the high tails of the p.d.f. in  Fig.~2.
\begin{figure}
\begin{center}
\mbox{\epsfig{file=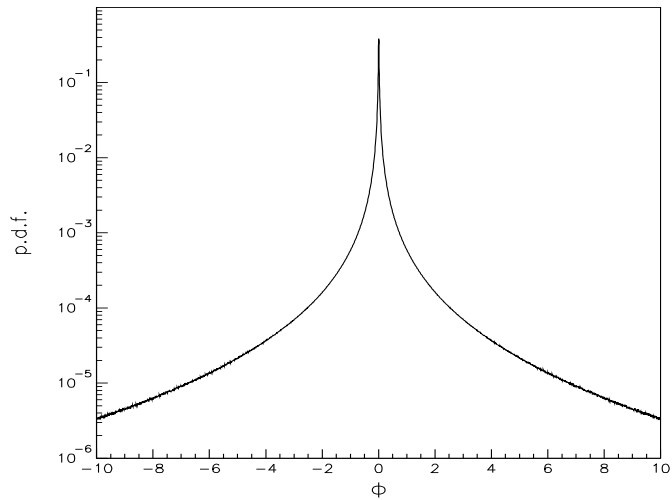,height=8cm,width=10cm}}
\end{center}
\vspace{-1cm}
\caption{The probability distribution
function (p.d.f.) of the gradient field $\phi$,
normalized to its r.m.s. value $\protect \sqrt{\langle\phi^2\rangle}$.} 
\label{fig2}
\end{figure}
\begin{figure}
\begin{center}
\mbox{\epsfig{file=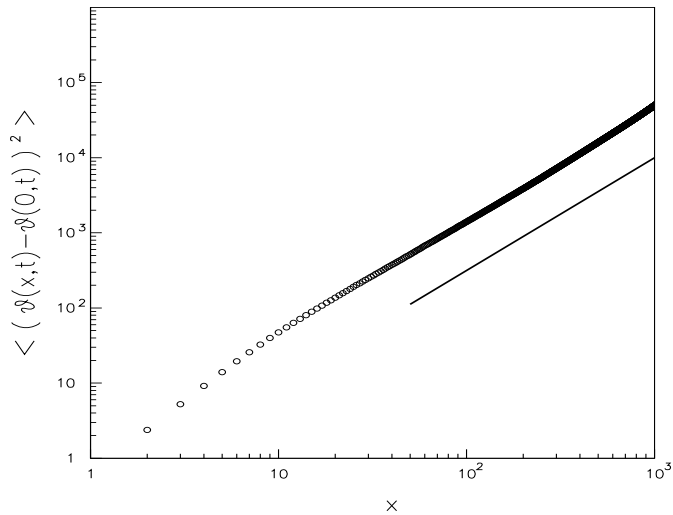,height=8cm,width=10cm}}
\end{center}
\vspace{-1cm}
\caption{The measured
second-order structure function for $\xi=0.5$. 
The solid line has the slope $2-\xi = 1.5$.}
\label{fig3}
\end{figure}

Let us now derive the predictions for the scaling exponents. The 
$\delta$-correlation in time ensures that correlation functions obey 
closed equation of motion. 
By using Gaussian
integration by parts, it is for example easy to derive the equation for 
the second-order correlation $C_2(x,t)=
\langle\phi(x,t)\,\phi(0,t)\rangle$\,:
\be
\label{corr2}
\partial_t C_2 = {d^2\over dx^2}\left[\left(2\kappa+S\right)
C_2\right] + F_L .
\ee
The solution is even and satisfies $\int C_2=0$.
In the inertial range of scales it decays
with the negative exponent $-\xi$.  The scaling of the
second-order structure function of $\theta$ is obtained by simple
integration. For $\xi<1$, its inertial scaling is 
$2-\xi$, as in Kraichnan model.

Let us now consider the scaling of higher-order correlation functions.
Closed equations of motion can be written both for the correlations
$C_{2n}=\langle \phi(x_1)\ldots\phi(x_{2n})\rangle$ and
those of the integrated field $\theta$.
Anomalous scaling is associated with the zero modes  
of the homogeneous equations.
The relevant equations for the zero mode $Z_{2n}$ of the 
$2n$-th order $\phi$ correlator are
\be
\label{operator}
{\cal H} Z_{2n}=\sum_{k\neq j, 1}^{2n}\nabla_j\,\nabla_k 
\left[S(x_{jk})\,Z_{2n} \right] = 0 .
\ee
where $x_{jk}=x_j-x_k$ and
$\nabla_j$ stands for the spatial derivative with respect to $x_j$.
Following Refs.~\cite{GK} and \cite{BGK}, we look for a scale-invariant 
solution of this equation as an expansion in powers of $\xi$.
Specifically, the structure functions are expanded as $S(x_{jk})=
D\left(1+\xi\,\log |x_{jk}| \right) + O(\xi^2)$ and
$Z_{2n}=Z_{2n}^{(0)}+\xi\,Z_{2n}^{(1)}+O(\xi^2)$.
The equations (for any $n$) at the lowest order in $\xi$ are trivially
satisfied by taking $Z_{2n}^{(0)}={\rm const.}=\bar{Z}_{2n}$. 
At the first order in $\xi$, the solution is 
\be
\label{perturba}
Z_{2n}\left(x_1,\ldots , x_{2n}\right)=
\bar{Z}_{2n}\left[ 1-{\xi\over 2} \sum_{k\neq j, 1}^{2n} 
\log |x_{jk}|\right].
\ee
The scaling exponent
(at order $\xi$) of the zero mode can now be obtained operating on 
$Z_{2n}$ with the Euler operator ${\cal E}=\sum_{j=1}^{2n}\left(x_j
\nabla_j\right)\,$.
It is easily checked that the equality ${\cal E} Z_{2n} = 
-\xi\,n(2n-1) \,Z_{2n}$ holds at the first order in $\xi$. 
The amplitude of the correction to normal scaling $-n\xi$ 
is thus $2n(n-1)\xi$.
This same value is obtained considering the equation for the
correlations of the integrated field $\theta$. 
Note that the anomalous factor $2n(2n-1)/2$ is simply related to symmetry
considerations, i.e. it is the number of couple of points in the correlation
function, as conjectured in Ref.~\cite{CF}.

\begin{figure}
\vfill \begin{minipage}{.495\linewidth}
\begin{center}
\mbox{\epsfig{figure=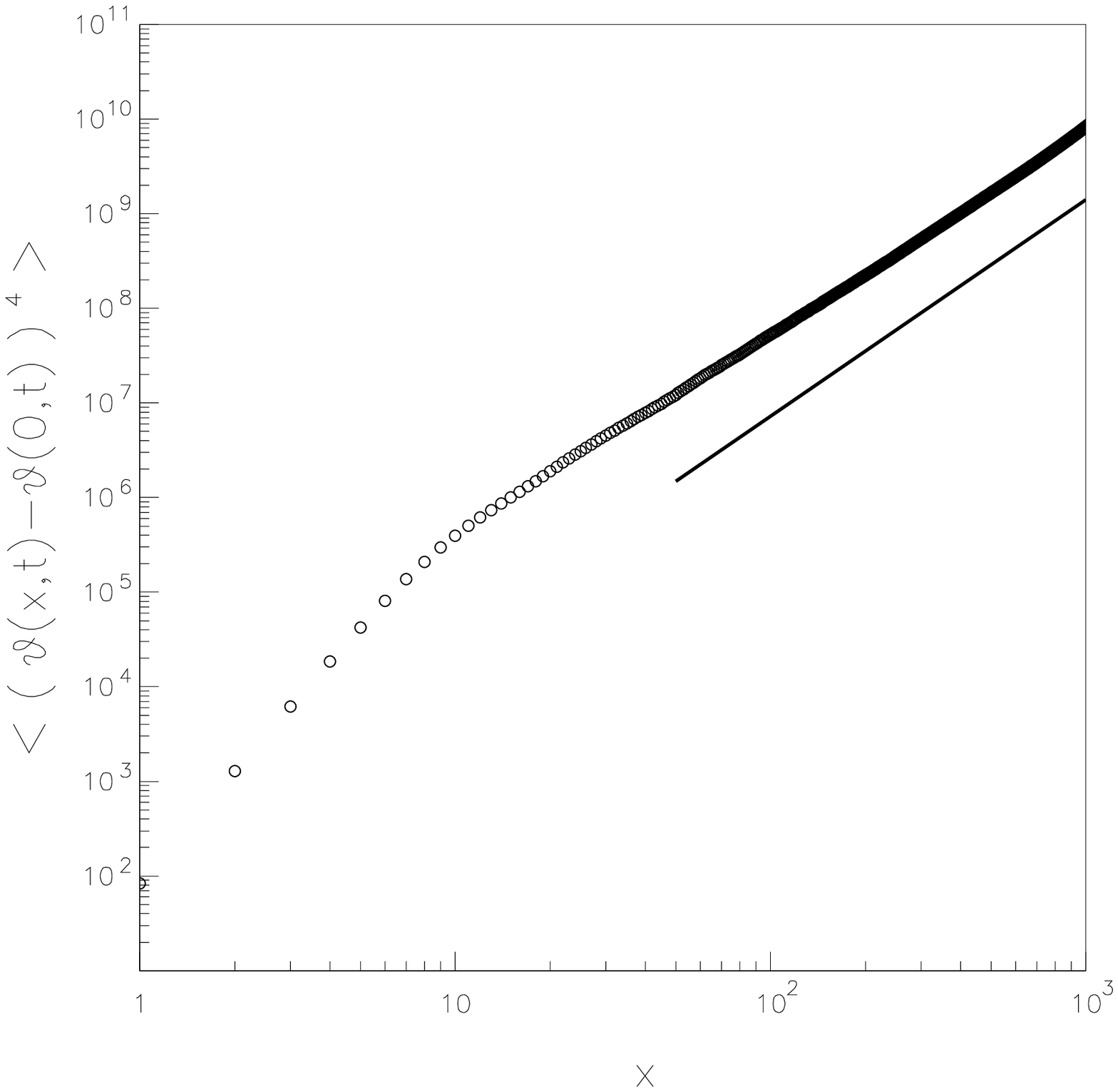,width=.9\linewidth}}
\end{center}
\end{minipage} \hfill
\begin{minipage}{.495\linewidth}
\begin{center}
\mbox{\epsfig{figure=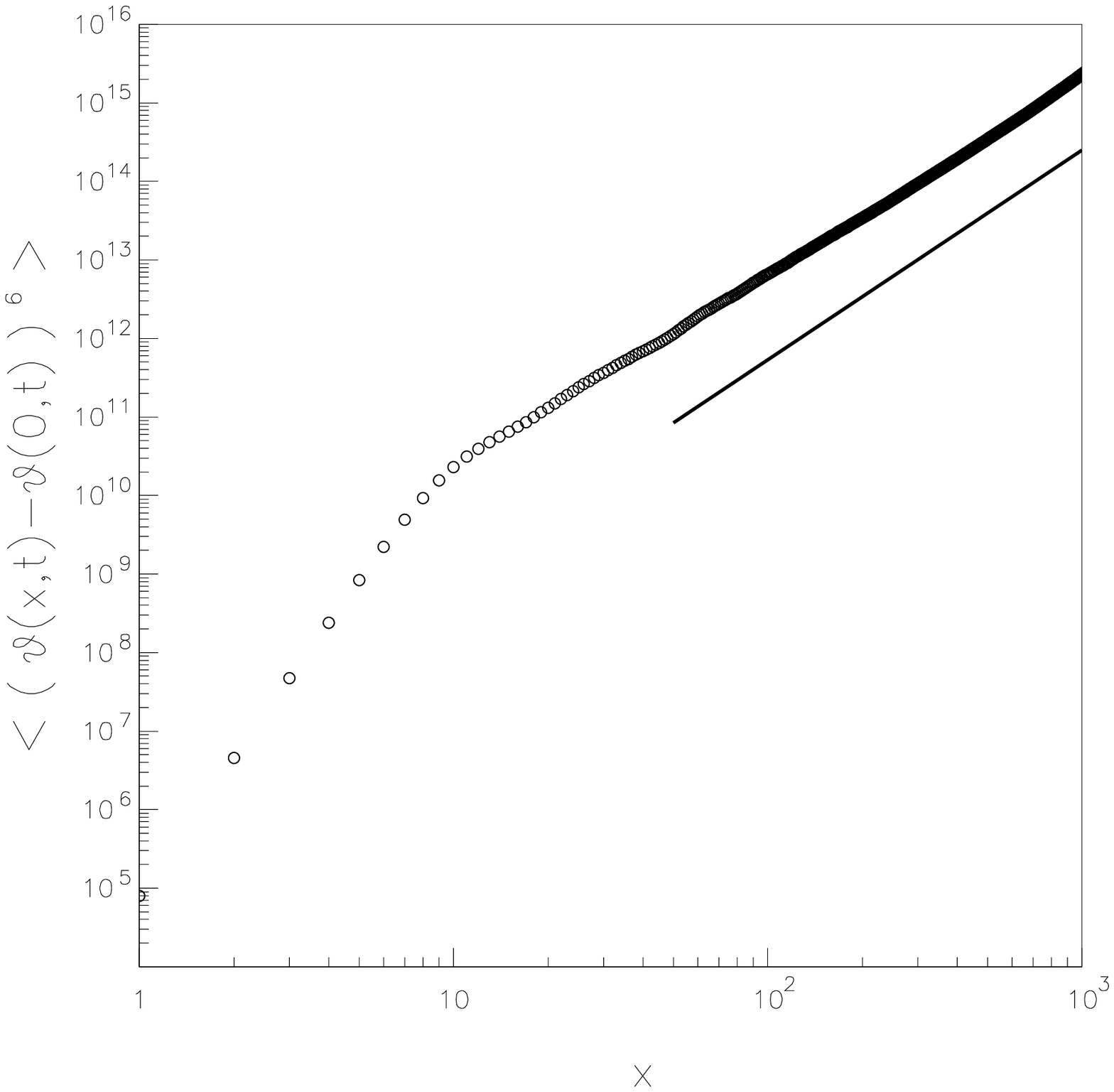,width=.9\linewidth}}
\end{center}
\end{minipage}
\label{fig4}
\vspace{-0.5cm}
\caption{The measured
fourth (left) and sixth (right) order structure functions for $\xi=0.5$. 
Solid lines have
the slopes predicted by (\protect\ref{predict}), i.e. $\zeta_4\simeq 2.29$ and 
$\zeta_6\simeq 2.67$.}
\end{figure}
For the comparison with numerical simulations, 
it is convenient to consider the scaling of the structure
functions $\langle | \theta(x,t)-\theta(0,t) |^{p}
\rangle \sim |x|^{\zeta_{p}}$.
In particular, the prediction for the fourth and the sixth order
are
\be
\label{predict}
\zeta_4=4-6\xi+O(\xi^2)\simeq{4\over 1+1.5\,\xi}\quad {\rm and}
\quad \zeta_6=6-15\xi+O(\xi^2)\simeq {6\over 1+2.5\,\xi},
\ee 
where we have used the classical procedure of Pad\'e approximants
\cite{ZJ}.
For the comparison we have considered $\xi=0.5$. The measured 
second-order structure function is presented in Fig.~3. 
The fourth and sixth
order structure functions are shown in Fig.~4.
It is worth to note that also low order moments have anomalous scaling.
The measured exponents for the moments 
of order $1/2$ and $1/8$ 
are $0.585$ and $0.163$. The corresponding normal values are $0.375$
and $\simeq 0.094$. It is an interesting issue the relation of
these anomalies with those of correlation functions. Note also that
$\xi=1$ separates the cases of negatively ($\xi<1$) and
positively ($\xi>1$) correlated increments of the velocity field.
We are currently investigating the influence of positive correlations 
on structures and anomalous scaling.

In conclusion, we have introduced and investigated a 1D white-in-time
passive scalar model. Numerical simulations show that the typical
configuration of the field is strongly structured.  The possible role
of structures as ``atoms'' for intermittency has recently been
addressed in Ref.~\cite{TD} for shell models (for white-in-time
shell models see Ref.~\cite{BBW}).  The conclusion of Ref.~\cite{TD}
was that the single structures are not quite elementary, i.e.
interactions between them and with the background play a key role for
scaling properties. This is also the case for our model. A naif
argument based on the scaling of a single structure would indeed
suggest the same asymptotic scaling as in Burgers equation,
i.e. a constant unit value.  Already the second order structure
function scales however with an exponent larger than one for $\xi <1$.
In the zero mode formalism used here,
structures and their correlations do not appear explicitly, but only
via their global statistical effects. Our results indicate that these
effects are correctly taken into account by perturbative expansions
around Gaussian limits. The agreement points in the direction of the
scale-invariant zero mode mechanism for the intermittency of the
model.

\bigskip
\noindent
{\bf Acknowledgments.} We are grateful to E.~Balkovski, B.~Dubrulle, 
G.~Falkovich, U.~Frisch, I.~Kolokolov and V.~Yakhot for helpful discussions.
We thank the ``Meteo--Hydrological Center of Liguria Region''
and the ``Swiss Scientific Computing Center'' where part of the numerical
analysis was done.

\end{document}